# Electromechanically Tunable Suspended Optical Nano-antenna

*Kai Chen[1 †\*], Gary Razinskas[1], Thorsten Feichtner[1,2,3], Swen Grossmann[1], Silke Christiansen[2,3], Bert Hecht[1\*]*

[1] Nano-Optics & Biophotonics Group, Experimental Physics 5, Wilhelm Conrad Röntgen-Center for Complex Material Systems (RCCM), University of Würzburg, Am Hubland, D-97074, Germany

[2] Christiansen Research Group, Max Planck Institute for the Science of Light, D-91058, Erlangen, Germany

[3] Institute Nano-Architectures for Energy Conversion, Helmholtz-Zentrum Berlin für Materialien und Energie GmbH, D-14109, Berlin, Germany

†Present address: International Center for Materials Nanoarchitectonics (MANA), National Institute for Materials Science (NIMS), Tsukuba, Ibaraki, 305-0044, Japan

[*]Chen.Kai@nims.go.jp; Hecht@physik.uni-wuerzburg.de

Coupling mechanical degrees of freedom with plasmonic resonances has potential applications in optomechanics, sensing, and active plasmonics. Here we demonstrate a suspended two-wire plasmonic nano-antenna acting like a nano-electrometer. The antenna wires are supported and electrically connected via thin leads without disturbing the antenna resonance. As a voltage is applied, equal charges are induced on both antenna wires. The resulting equilibrium between the repulsive Coulomb force and the restoring elastic bending force enables us to precisely control the gap size. As a result the resonance wavelength and the field enhancement of the suspended optical nano-antenna (SONA) can be reversibly tuned. Our experiments highlight the potential to realize large bandwidth optical nanoelectromechanical systems (NEMS).

Optical nano-antennas[1, 2] provide unprecedented capabilities to manipulate light-matter interactions at the nanoscale, enabling a large number of exciting applications, such as surface-enhanced spectroscopy, sensing and light harvesting.[3-9] While the plasmonic resonances of nano-antennas can be tailored by changing their size, shape and composition, it is particularly appealing to be able to dynamically control or modulate such resonances, thereby tuning their resonance wavelengths, scattering efficiencies, and emission characteristics. So far, tuning mechanisms have involved (electro)mechanical methods,[10-12] modulation of liquid crystals,[13-17] responsive molecules,[18-21] phase change materials,[22-24] and electrical modulation of graphene.[25-27] In this letter, we demonstrate a new type of reconfigurable optical nano-antenna operating in the visible spectral range. The nano-antennas are made from single-crystal gold flakes by means of focused-ion beam (FIB) milling. Due to the superior mechanical properties of this material, in contrast to previous work,[12] it is possible to suspend individual gold nanostructures in air without supporting material over a trench in the substrate. We show that by varying the applied voltage, the equilibrium between repulsive Coulomb forces and restoring elastic forces can be tuned reversibly. Therefore, we are able to control the antenna's optical properties by tuning its gap width. This type of reconfigurable optical antenna opens up plenty of opportunities in optomechanics, sensing, and active plasmonics.

The suspended two-wire nano-antenna's design is illustrated in Figure 1. The nano-antenna is suspended over a trench in a glass substrate. The two antenna wires/arms are physically and electrically connected to the same contact pad (Figure 1a, top) through two thin, flexible and highly conductive wires (leads). A second contact pad at the opposite side of the trench serves as counter electrode. Both the nano-antenna and the two leads are fabricated from the same single-crystalline gold flake ensuring excellent conductivity, mechanical properties, as well as a pronounced plasmonic response,[28] undisturbed by extended lattice defects and grain boundaries typical for vapor-phase deposited gold films.[29] As shown by Prangsma *et al.*,[30] there exists an optimal location for connecting the leads to the nano-antenna arms with marginal influence on its fundamental plasmon resonance. As illustrated in Figure 1a, applying a voltage induces accumulation of positive charges on the nano-antenna arms and leads, as well as a corresponding quantity of negative charges on the counter electrode. The equal charges on the antenna wires result in a repulsive Coulomb force

$$F_c = k_c \frac{Q^2}{d^2}$$

where $Q$ is the effective charge on the antenna arm (including the lead), $d$ is the distance between the "center of charges" defined in analogy to a center of mass, and $k_c=1/4\pi\varepsilon_0$. The quantity of the induced charges, and hence the magnitude of $F_C$, depends on the applied voltage and the distance between the SONA and the counter electrode. As in an electrometer, $F_C$ pushes the two antenna arms apart and thereby increases the nano-antenna gap until it is balanced by the restoring elastic force $F_E$ exerted by the bending leads

$$F_c = F_E$$

$$F_E = k_e \Delta d$$

where $k_e$ is the elastic constant of the leads and $\Delta d$ is the effective deflection of the leads.

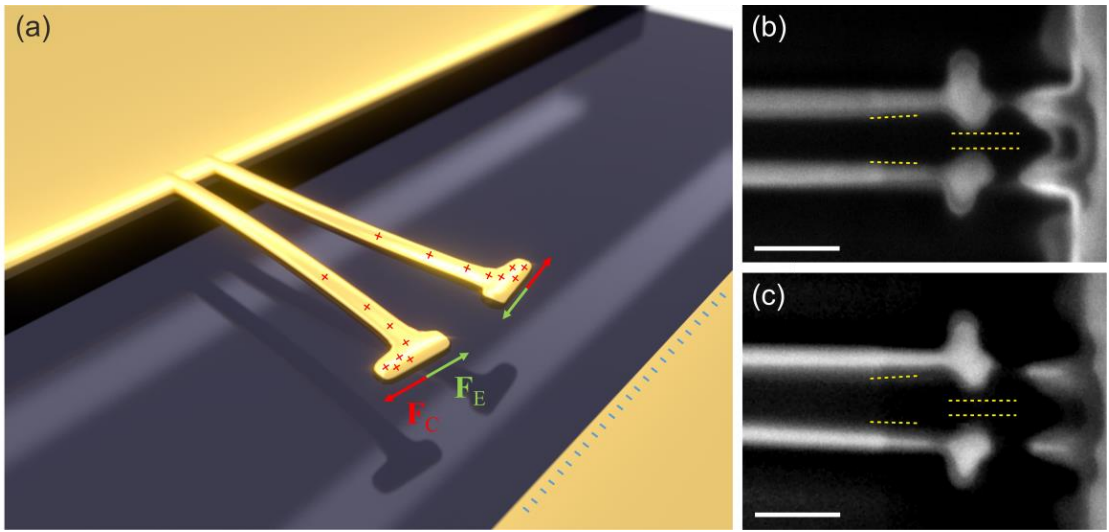

**Figure 1**. (a) Artistic view of the suspended optical nano-antenna: The gap width is determined by the equilibrium of the Coulomb force $F_C$ and the restoring elastic force $F_E$ (deflection not to scale). Scanning electron microscopy (SEM) images directly visualize the active tuning of the antenna gap width of a fabricated SONA from (b) 40 nm at 0 V to (c) 70 nm at 20 V applied voltage. The two pairs of yellow dashed lines indicate the gap changes and serve as guides to the eye. The scale bars are 200 nm in both panel b and c.

We first directly visualized the operating principle of the proposed SONA inside an SEM. A pronounced gap widening was observed when a voltage was applied. The gap of the antenna increases from 40 nm at 0 V (Figure 1b) to 70 nm at 20 V (Figure 1c). In Figure 1b two pairs of the yellow dashed lines mark the width of the gap and the distance between the leads, respectively. The same lines are reproduced in Figure 1c indicating the widening of the antenna gap. Such observations inside SEM directly confirmed the proposed working principle of the SONA and allowed us to verify its repeatability.

To determine the tunability of the optical properties of the SONA, we use white light dark-field scattering spectroscopy. For this purpose the SONA design needed to be optimized towards larger antenna-counter electrode distance in order to sufficiently separate the light scattering signals of the SONA from those of the counter electrode rim (cf. Figure 2a). The nano-antenna investigated here is slightly asymmetric with arm lengths of 160 nm and 190 nm. The top antenna arm features a small sharp bulge due to imperfect FIB milling. The presence of such small features has negligible effect on the optical properites of the antenna and does not degrade its resonance.[30] In the optimized design, the antenna has a distance of ~ 645 nm to the counter electrode (Figure 2a) and the leads have a length of 1700 nm. The top (bottom) lead has a width of about 65(75) nm. The antenna is suspended roughly 300 nm above the bottom of the trench. Evaluation of SEM images of the SONA obtained under different viewing angles as shown in Figure 2a yields the 3D arrangement of the structure before applying a voltage. The two antenna arms are almost in the same horizontal plane with the top arm bending upwards a little bit. We found that such bending occurs due to the stress induced by ion implantation during the FIB milling process. This initial 3D arrangement of the antenna has been taken into account in the following numerical simulations. Since the stress-induced bending breaks the symmetry, mechanical degrees of freedom both parallel and perpendicular to the substrate need to be considered.

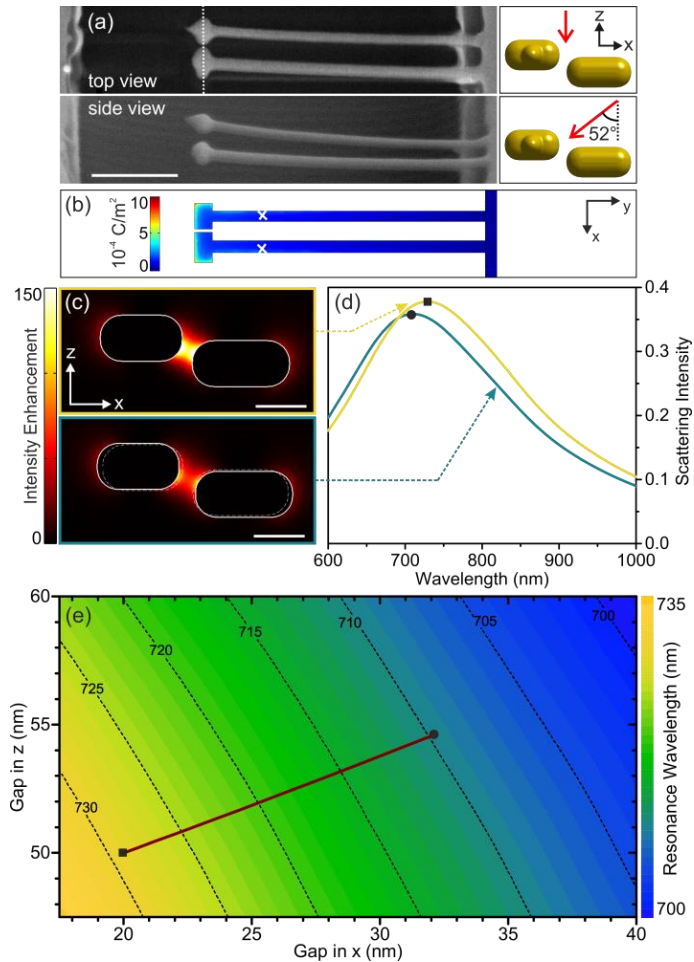

**Figure 2.** Suspended optical nano-antenna for electromechanical tuning of plasmon resonance. (a) Top-view and side-view SEM images of the fabricated SONA. The respective observation angles are indicated by the red arrows in the pictorial representations to the right. (b) COMSOL simulated charge density distribution on the antenna and the lead wires for an applied voltage of 10 V. The white cross sign "×" indicates the position of the effective center of charges. (c) FDTD simulated near-field intensity distributions of the SONA with initial gap size obtained from SEM images (top) and increased gap size (bottom) and (d) the corresponding simulated scattering spectra. The near-field intensity distributions taken along the cross-section indicated by the white dotted line in panel (a) are obtained at their respective resonance wavelengths. (e) FDTD simulated antenna resonance wavelengths as a function of the gap projections in *x* (parallel to the substrate) and *z* axis directions (perpendicular to the substrate). The resonance wavelengths are color-coded according to the color bar to the right. The solid line indicates the trajectory used in our analytical model. The scale bars in panels (a) and (c) are 500 nm and 100 nm, respectively.

To understand the voltage dependence of the antenna's equilibrium configuration, it is important to know the charge distribution on the antenna and the lead wires. We therefore performed electrostatics simulations of the structure using COMSOL as a function of the antenna gap (simplified rectangular geometry, see Supporting Information, Part I). Figure 2b shows the simulated surface charge density distribution at an applied voltage of 10 V. As expected, the charge density is the highest on the antenna arms, especially at the outer corners. The charges are drawn to the antenna arms by the counter electrode and pushed to the outer corners due to the repulsive Coulomb force between the charges on the two arms. It is noted, however, that the induced charges are distributed over the entire surface of the antenna as well as the lead wires. While the charge density amplitude is the highest on the antenna arms, the center-of-charge position, needed to calculate $F_C$, is also influenced by the size of the area over which the charges are distributed. Indeed, we find that the center of charges (see also Supporting Information Figure S2) is located on the lead wires, ~400 nm away from the antenna, as indicated by the white "×" in Figure 2b.

The optical properties of the SONA were simulated using the finite-difference time-domain (FDTD) method. The possibility of out-of-plane bending of the antenna arms is taken into account by considering projections of the antenna gap in *x* (parallel to the substrate) and *z* direction (perpendicular to the substrate). The initial out-of-plane offset between the two antenna arms was determined to be 50 nm (20 nm) in *z*-(*x*-) direction based on the SEM images in Figure 2a. Figure 2c shows near-field intensity profiles of the SONA without (top) and with (bottom) the applied voltage recorded at the respective resonance wavelengths. The simulated scattering spectra of the same structures are shown in Figure 2d. For the zero voltage case the SONA shows a peak at 729 nm (yellow curve). For an applied voltage of 40V, the *x*-projection of the gap size increases to 32 nm and the *z*-projection to 55 nm, which causes the plasmon peak to blue-shift to 710 nm while the peak intensity diminishes (cyan curve) slightly. This is attributed to the decreasing coupling between the two antenna arms as indicated by the near-field intensity profiles in Figure 2c with and without applied voltage. Figure 2e shows a 2D map of the antenna resonance as a function of the *x*- and *z*- projections of the nanogap obtained by FDTD simulations. In the experiments, the antenna gap traces a trajectory that must be a straight line in this map. The most likely trajectory of the antenna gap change corresponds to the superimposed red line, which is obtained from the analytical model we developed (Supporting Information, Part I), where the system is modeled as two conductive cantilevers that undergo elastic bending. By solving the equations for the force equilibrium of the two cantilevers (leads), we are able to calculate the antenna gap size at each voltage value. A subsequent FDTD simulation leads to the resonance wavelength at that antenna gap (or voltage) corresponding to a point on the red line in Figure 2e. It is noted that the starting point of the trajectory is extracted from the SEM images (Figure 2a) and is therefore predetermined. The slope of the trajectory is also fixed because it is determined by the leads' moments of inertia in $x$ and $z$ directions, which are fixed once the dimensions and Young's modulus of the leads are fixed (Supporting Information, Part I). The only fitting parameter whithin this model is the absolute amount of charges on the system as a function of the applied voltage (corresponding to the system's capacitance), which determines the end point of the deflection trajectory.

To demonstrate dynamic tuning of the SONA's optical response, we ramped up the voltage at a rate of 1V/3s from 0 to 40 V and subsequently ramped it down to zero at the same pace. Figure 3a shows white-light scattering spectra of the nano-antenna recorded for fixed discrete voltages during the voltage sweeps. The minimum and maximum applied voltages are indicated on the respective spectra. In the beginning, the SONA exhibits a plasmon resonance at 730 nm. As the voltage increases, the resonance peak continuously shifts to shorter wavelength. The blue-shift reaches its maximum at a peak wavelength of 713 nm with an applied voltage of 40 V. Afterwards, as the voltage is reduced back to 0 V, the resonance returns to its original spectral position. This clearly demonstrates the reversible tunablity of the SONA. We also note that the intensity of the plasmon peak continuously decreases as the voltage increases. The peak intensity drops by ~25% when the voltage reaches 40 V. This is consistent with the FDTD simulations (Figure 2d) and the expectation that for wider antenna gaps the coupling between the antenna arms is weaker and hence a smaller scattering cross-section of the bonding mode of the two-wire nano-antenna should be observed.[2] Again, the peak intensity fully recovers as the voltage is reduced to zero. To further confirm the electromechanically tunable characteristic of the SONA, we compared the scattered light intensity of the antenna with that of the counter electrode rim using the images formed by the scattered light on the CCD of the spectrometer (see details in Supporting Information, Part II). We observed that the former varied accordingly with the voltage, decreasing and increasing with voltage ramping up and down, while the latter stayed unchanged during the sweeping. This excludes spectral changes caused by possible instability of the instruments and confirms the reconfigurablity of the SONA.

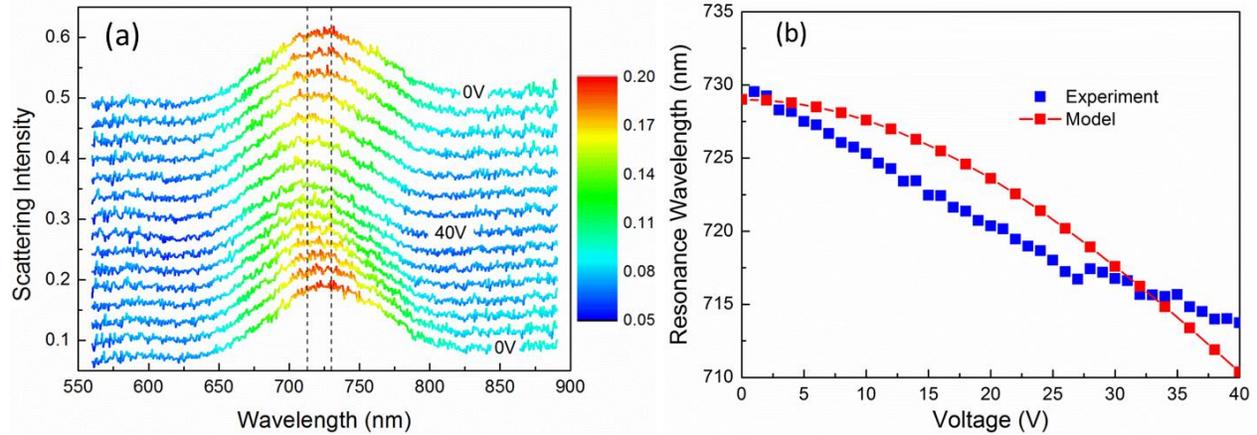

**Figure 3.** (a) Experimental demonstration of the reversible tunability of the SONA. The scattering spectra are color-coded according to the color bar on the right. The vertical dashed lines denote the resonance peaks at 0 and 40 V. (b) The relationship between the plasmon resonance wavelength and the applied voltage obtained from the experiments (blue) and the analytical model (red, Part I, Supporting Information).

Figure 3b shows the relationship between the resonance wavelength and the applied voltage ($\lambda \sim U$) obtained from the experimental data as well as from the analytical model (Supporting Information, Part I). We note that the experimental data display a near-linear relationship while the model data seem to suggest a nonlinear (parabolic) behavior. Parabolic behavior of the displacement with increasing voltage is also expected when considering an analytical model for the displacement as voltage enters the equations inside the square terms. The exact causes behind the observed discrepancy are so far not known to us. It is likely a combination of multiple factors, such as the modified modulus of gold due to ion implantation[31] and leakage currents in the substrate. Nonetheless, with system's capacitance as the only free parameter, the model yields results that within small error margins quantitatively describe the SONA's spectral shift as a function of applied voltage.

In conclusion, we have demonstrated a new nano-opto-mechanical device, i.e. a suspended resonant optical nano-antenna fabricated from single crystalline gold flakes. Reversible tuning of the antenna gap, and thus the antenna resonance, was achieved by applying a voltage that induces a repelling electrostatic force between the antenna arms. It is anticipated that the two-wire nano-antennas or other types of SONAs can find applications in a variety of fields, such as active plasmonics,[32-34] plasmonic optomechanics[35, 36] and plasmonic optical forces.[37, 38] This new type of nano-antenna provides an excellent platform to investigate the intriguing combination of plasmonics and nanoelectromechanical systems (NEMS).

## Methods

*Au Flake Synthesis*. Gold flakes were synthesized following the recipe in reference with small minor revisions.[39] In brief, a 100 mM $HAuCl_4$ aqueous solution was added to ethylene glycol solution followed by subsequent heating at 90°C in an oven. Gold flakes with various sizes were formed in the solution after 12 hours. Then the gold flakes were rinsed with ethanol for storage.

*SONA Fabrication*. Large gold contact pads were fabricated by photolithography followed by milling the trenches inside a commercial FIB dual-beam machine (Helios Nanolab, FEI Company) with gallium (Ga) ions accelerated with 30 kV. The Au flake solution was drop-casted onto the samples and large Au flakes were selected and pushed over the trenches with a micromanipulator (Suss MicroTec AG). The Au flakes were large enough to cover part of the contact pads leading to an ohmic contact between them. Finally the SONA were fabricated with FIB milling.

*In-situ SEM Characterization*. The SEM characterization of the SONA was carried out inside a commercial SEM (Ultra plus, Carl Zeiss Microscopy). Two beryllium copper (BeCu) probes attached to two in-situ micromanipulators (Kleindieck) were used to make the electrical contact with the gold contact pads. A sourcemeter (Keithley 2601A) was used to apply the voltage as well as monitor the current through the system.

The external voltage was applied onto the structure in an off-and-on manner as a proof-of-concept experiment. For each value, the applied voltage directly increased from zero to target voltage value, and stayed there for a few minutes allowing SEM image-taking before dropped to zero instantly. This procedure was repeated for several voltage values to observe the corresponding changes of the antenna configuration.

*Dark-field Scattering Spectroscopy*. White light scattering spectroscopy was performed on a home-build system. More details can be found in Ref. 21. In brief, a collimated white light beam was focused onto the sample surface from below with grazing angle incidence. The light was polarized along the longitudinal axis of the antenna. The scattered light from the antenna was collected and fed to an EM-CCD of a commercial spectrometer (Andor Technology Ltd.) with integration time of 1s. The scattering was calculated with reference to the incident white light on the trench. Again, a sourcemeter (Keithley 2601A) was used to apply the voltage as well as to measure the current. In this case, the voltage was increased or decreased in a step-wise fashion and the corresponding spectra were recorded.

*Simulation*. Finite-difference time-domain (FDTD) simulations were performed using commercial software (FDTD Solutions v8.7.1, Lumerical Solutions Inc.). The electrostatic multiphysics simulations of the antennas were done in COMSOL 4.4.

## Acknowledgement

The authors would like to thank Monika Emmerling and Martin Kamp for help with FIB nanofabrication. The authors would also like to thank Johannes Kern and René Kullock for helpful discussions. K.C. acknowledges the financial support from Alexander von Humboldt Foundation during his research stay in Germany. T.F. and S.C. gratefully acknowledge financial support from the European Commission within the FP7 project UNIVSEM (grant agreement number 280566).

# Supporting Information

## Electromechanically Tunable Suspended Optical Nano-antenna

Kai Chen[1,†,*], Gary Razinskas[1], Thorsten Feichtner[1,2,3], Swen Grossmann[1], Silke Christiansen[2,3], Bert Hecht[1,*]

[1] Nano-Optics & Biophotonics Group, Experimental Physics 5, Wilhelm Conrad Röntgen-Center for Complex Material Systems (RCCM), University of Würzburg, Am Hubland, D-97074, Germany

[2] Christiansen Research Group, Max Planck Institute for the Science of Light, Günther-Scharowsky-Str. 1 / Bldg 24, D-91058 Erlangen, Germany

[3] Institute Nano-Architectures for Energy Conversion, Helmholtz-Zentrum Berlin für Materialien und Energie GmbH, Hahn-Meitner-Platz 1, D-14109 Berlin, Germany

†Present address: International Center for Materials Nanoarchitectonics (MANA), National Institute for Materials Science (NIMS), Tsukuba, Ibaraki, 305-0044, Japan

*Chen.Kai@nims.go.jp; Hecht@physik.uni-wuerzburg.de

## I. Analytical model for antenna lead deflection

### *1. Basic setup*

For the analytical calculation of the lead beam bending, we used a simplified geometry with rectangular cross section (see Figure S1) and the following dimensions: The overall length of the two leads **1** and **2** is $L = 1700$ nm along the *y*-direction; The height *h* in *z*-direction is 50 nm and the width of the two leads is $w_1 = 65$ nm and $w_2 = 75$ nm, respectively. As shown in Figure 2 in the main text, the antenna arms exhibit a slight position offset. We model this effect assuming a completely straight lead beam connected to the second antenna arm that is displaced in *z*-direction by 50 nm (see Figure S1b). The 50 nm offset is chosen based on examination of Figure 2a.

We further simplify our analytical model by assuming that the induced charges ($Q_{1/2}$) are located at two points, called center of charges, $\boldsymbol{r}_{c1/c2}$. The Coulomb force will act between the center of charges and a corresponding effective length $L_{eff}$ of the lead is used in the model (see Figure S1c). The lead will bend from its base at the electrode until $\boldsymbol{r}_c$ and the tilting angle at this point is linearly interpolated to the lead end to obtain the changes in the antenna gap size.

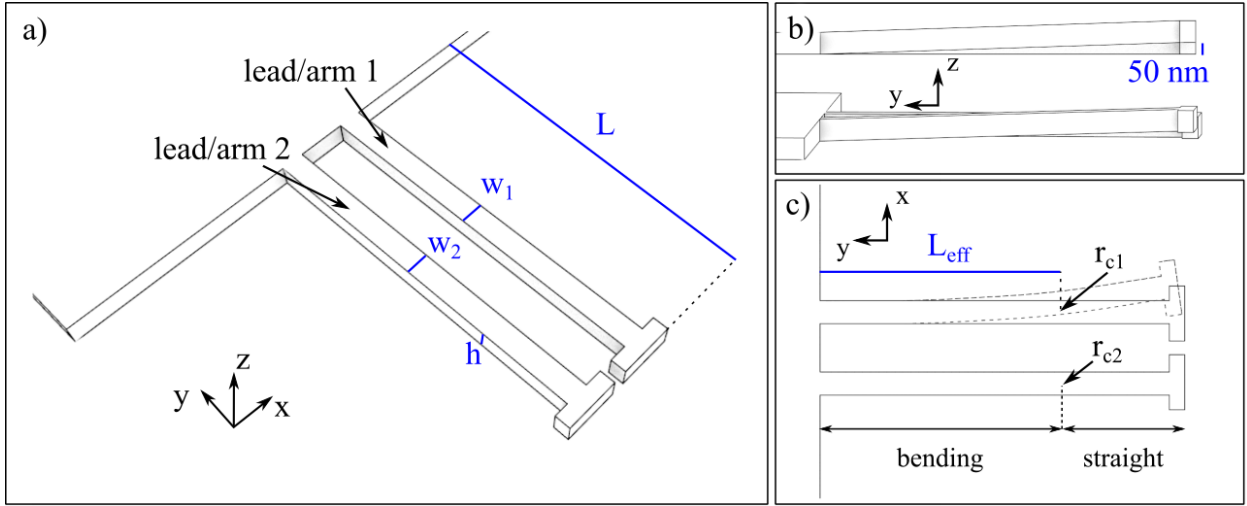

**Figure S1**. Geometry for the analytical model of the lead beam bending. (a) General setup with two leads of identical length $L$ and height $h$, but different width $w_1$ and $w_2$. (b) Offset of antenna arm **2** in $z$-direction in side view (top) and perspective view (bottom). (c) The charges on the antenna and the lead are treated as localized at the center of charges $\boldsymbol{r}_{c1/c2}$. An effective length $L_{eff}$ is thus used for the bending calculation. The lead wires are bending until this point and straight for the remaining part with a fixed angle (top view).

*2. Center of charges*

The position of the center of charges is calculated via weighted integrals over the surface charge density σ, which is obtained by electrostatics simulations performed in COMSOL 4.4 (see Figure S2b). As an example, the center-of-charge coordinate in $x$-direction is given by:

$$x_c = \frac{\int \sigma(\boldsymbol{r}) \cdot x(\boldsymbol{r}) dA}{\int \sigma(\boldsymbol{r}) dA}$$

, where $A$ is the surface of the lead wire and the antenna. These calculations also justify the approximation of equal charge accumulation on the two leads.

If we assign an effective capacitance $C$ to the system, the induced charges on each lead can then be expressed as $Q_{1/2} = \frac{1}{2} C \cdot U$, as a function of the applied voltage $U$.

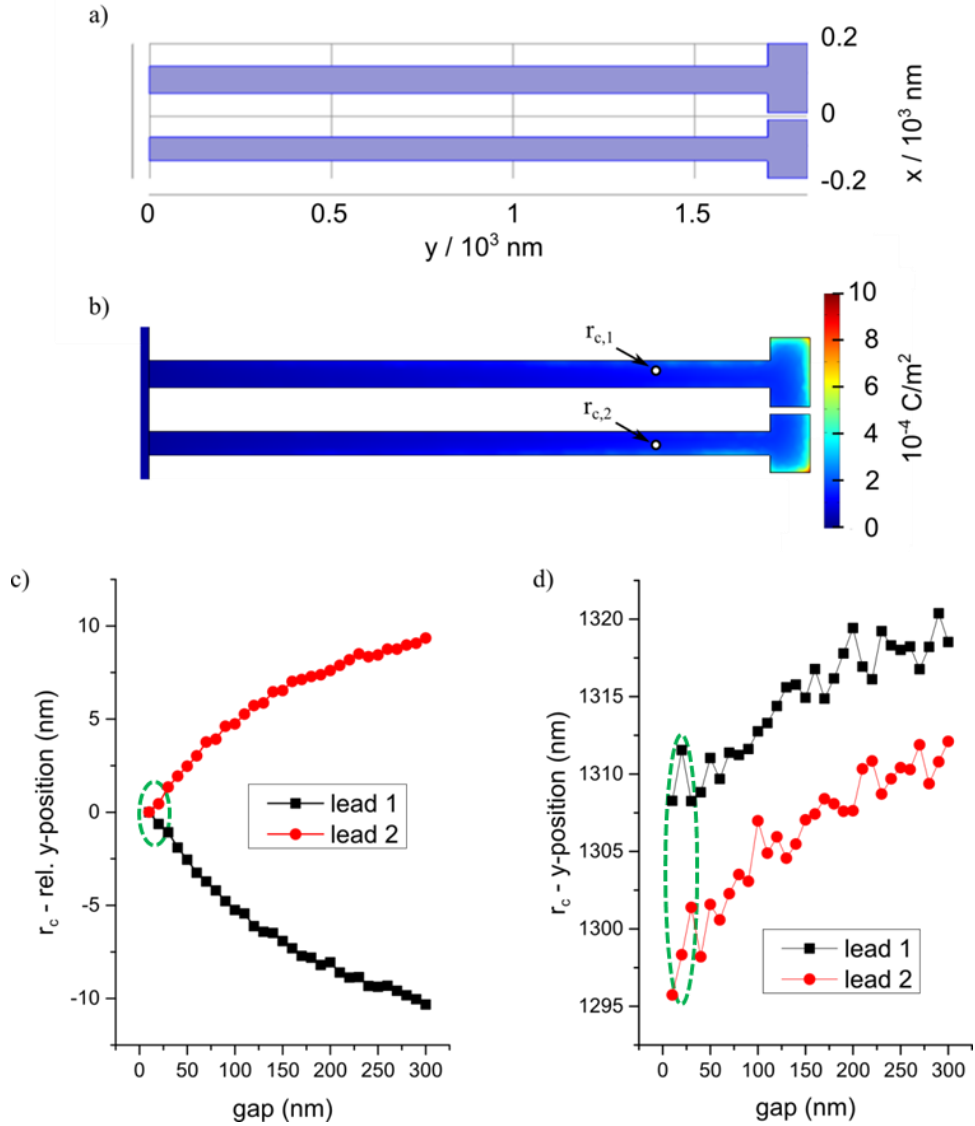

**Figure S2**. COMSOL calculation of the charge distribution on the SONA. (a) Part of the simulated geometry. The lead wires connect to the contact pads (not depicted here) at y = 0 nm. (b) Surface charge density σ on the SONA structure with an applied voltage of 10 V and nanogap size of 20 nm. The calculated position of the center of charges $r_c$ is indicated on each lead wire. (c) The relative position changes (in x-direction) of the centers of charges $r_c$ on the leads as a function of the antenna's nanogap size. The numbers on y-axis give the changes of $r_c$ relative to the initial state with 20 nm nanogap. (d) The position changes (in y-direction) of the centers of charges $r_c$ on the leads as a function of the antenna's nanogap size. The green dashed ovals in panels c and d indicate the working regime in our experiments.

The centers of charges on the two leads, as indicated in Figure S2b, are located $L_{eff} \sim 1305$ nm from the lead base and have a starting separation $\Delta x_0 \sim 219$ nm (see Figures S2b). Assuming a linear increase of the relative z-offset along the lead, we have, at the position of the centers of charges, $\Delta z_0 = 50$ nm $\cdot \frac{L_{eff}}{L} \sim 38.2$ nm, where 50 nm is the estimated z-offset of the two antenna arms. Figure S2c shows the *relative* position changes of the centers of charges in *x*-direction as a function of the gap size. It is noted that in Figure S2c nanogap sizes up to 300 nm were used in the calculation. However, in the experiments, the actual nanogap size is below 40 nm and the corresponding relative position changes of the centers of charges are negligible. Therefore, we fixed the *relative* positions of the center of charges on the leads in our model. Figure S2d shows the *absolute* position changes of the center of charges in *y*-direction as a function of the gap size. As the gap widens, the changes in *y*-direction are rather small, which is expected as the nanogap is oriented along the *x*-direction. Therefore, in our model, the y-coordinates of $r_c$ are neglected and all the calculations will be performed in the *x-z*-plane (i.e. $r_0 = (x_0, z_0)$).

The initial distance vector between the two repelling centers of charges therefore reads as $\Delta r_0 = (219, 38.2)$ nm. For a given applied voltage $U$ we can set up two vector equations describing the force equilibrium for the two leads, leading to four scalar equations for the four unknown displacements: $\Delta x_{Leff,1}, \Delta x_{Leff,2}, \Delta z_{Leff,1}$ and $\Delta z_{Leff,2}$. The solution to this set of equations is then used to derive the change of the width of the antenna's nanogap. From this we then obtain the relationship between the applied voltage $U$ and the antenna gap, which is directly linked to the antenna resonance wavelengths through FDTD simulations as shown below. Finally, we obtain the relationship between the resonance wavelength of the antenna and the applied voltage and compare it with the experiment.

### 3. Resonance - voltage relationship

The lead deflection for a given applied voltage depends on the equilibrium of two forces: the repelling Coulomb force $F_C$ and the restoring elastic force $F_E$:

$$F_C = \frac{1}{4\pi\varepsilon_0} \frac{Q_1 \cdot Q_2}{r^2} \frac{\boldsymbol{r}}{|\boldsymbol{r}|} = F_E = \frac{3E_{Au}I}{L_{eff}^3} \Delta \boldsymbol{r}$$

Here, $\varepsilon_0$ is the vacuum permittivity, $\boldsymbol{r} = \pm(\boldsymbol{r}_{c1} - \boldsymbol{r}_{c2})$ the distance vector between the two centers of charges, $E_{Au} = 79$ GPa the Young's modulus of gold, $I$ the moment of inertia of the lead, $L_{eff}$ the length of the lead beam up to the center of charges, and $\Delta \boldsymbol{r}$ the beam deflection at the center of charges. The moment of inertia $I$ for the two leads, both with rectangular cross section, is defined as:

$$\boldsymbol{I} = \begin{pmatrix} \frac{hw_{1,2}^3}{12} \\ \frac{h^3 w_{1,2}}{12} \end{pmatrix}.$$

The equilibrium equations were solved in Mathematica (version 10) and gave rise to multiple solutions. We chose the real solution that describes leads bending away from each other. To retrieve the bending at the end of the leads, where the antenna arms are located, we calculate the bending angle at the center of charges[1]:

$$\theta = \left(\frac{FL^2}{EI}\right)$$

The movement of the antenna arms can then be obtained via simple linear interpolation. For example, in $x$-direction, it is given by

$$\Delta x_L = \Delta x_{L_{eff}} + \theta \cdot (L - L_{eff})$$

Finally we are able to obtain the bending of the antenna arms at different applied voltages with the system capacitance $C$ as the only free parameter. Figure S3 shows the bending amplitude at the centers of charges and the antenna arms with $C = 3.0 \times 10^{-18}$ F showing a small bending in the beginning followed by a nonlinear increase for larger voltages.

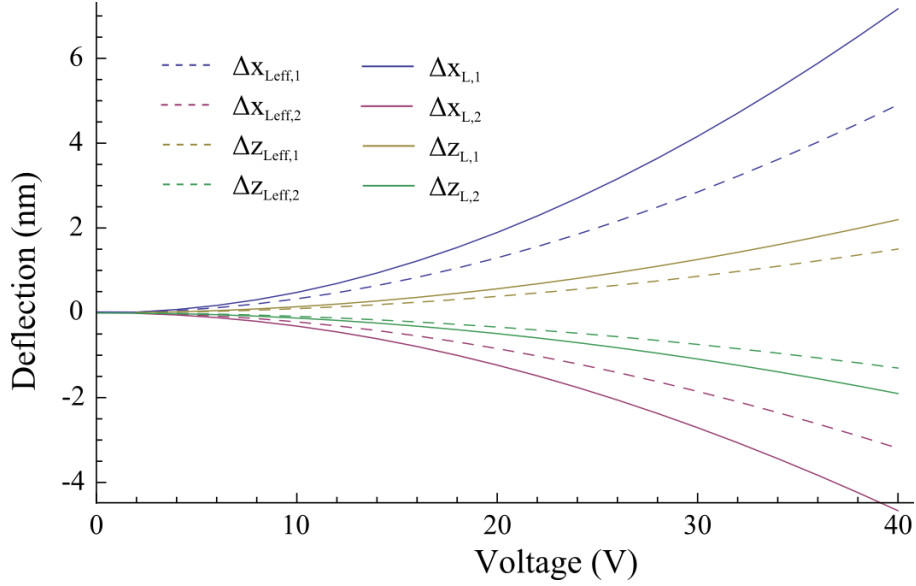

**Figure S3**. The bending of the two leads at the centers of charges (dashed lines) and at the ends (solid lines) under different applied voltages.

Using an interpolating function we can translate the antenna arm bending to the antenna gap changes via $x_{gap} = x_{gap,0} + \Delta x_{L,1} - \Delta x_{L,2}$ and $z_{gap} = z_{gap,0} + \Delta z_{L,1} - \Delta z_{L,2}$, where $x_{gap,0}$ and $z_{gap,0}$ are the initial gap size in $x$ and $z$ direction, respectively. So far we have established the relationship between the applied voltage ($U$) and the antenna nanogap from this analytical model. However, in the experiments, we obtained resonance wavelength *vs* voltage ($\lambda$-$U$). Thus to compare the model with the experimental results, we also need the relationship between the antenna nanogap and the resonance wavelength, which can be obtained through FDTD numerical simulations.

The SONA shown in Figure 2a was realistically modeled in FDTD according to the SEM images taking rounded edges, corners, and small features such as the small bulge on the top arm into account. The wavelength-dependent dielectric function of gold was modeled by an analytical fit to experimental data.[2, 3] The obtained data between the plasmon resonance and the nanogap was then interpolated based on a third-order polynomial fit.

At last, using the nanogap as a link, we obtain the relationship between the antenna resonance and the applied voltage and compare it to the experimental results. We vary the only free parameter, the capacitance $C$ of the system, to achieve the best agreement between experiment and model. The result is depicted in Figure 3b in the main text with $C = 3.0\times10^{-18}$ F. The analytical model does not replicate the linear dependence of the antenna resonance on the applied voltage as observed in the experiment. However, with the system capacitance as the only free parameter, the results from the model are on the same order as the experimental data. Several factors can contribute to the discrepancy between the model and the experiment, such as the Young's modulus of the single-crystal gold. FIB milling on suspended gold flakes may lead to changes to the mechanical properties of the gold lead wires.

## II. CCD imaging of the scattered light from the SONA

To exclude any possibility that the observed spectral change might come from the instablity of the sample or other instrument artifacts, we repeated the dynamic tuning experiment shown in Figure 3a and recorded the scattered light images instead using the same CCD camera of the spectrometer. Figure S4a shows the collected image with no voltage applied. The intensity of each pixel is color-coded according to the color bar on the right. The bright spot enclosed by a black dashed line (7×7 pixel) is formed by the scattered light from the SONA while several other bright spots on the top part of the image are formed by the undesirable scattered light from the milled trench edge (left part in Figure 2a), which has small features that can scatter light. Nonetheless, these bright spots actually provide us excellent reference points because applying a voltage should not affect the scattered light from the edge. One of them indicated by a red dashed line is then used as a reference.

In each image, we sum up the light intensity from all the pixels in either box and plot them against the frame number as shown in Figure S4b. The intensity from the SONA (black solid squares) gradually decreases with increasing voltage until a maximum voltage of 40V is applied. Afterwards, the intensity progressively goes up and recovers back to the original level as the voltage goes to zero. An intensity change up to 20% is observed, which is in good agreement with the scattering spectra as shown in Figure 3a. By contrast, the intensity from the reference point (red solid squares) stays on the same level with a slight 1.5-1.7% fluctuation. We thus confirm the reversible change of the scattering spectra of the SONA stems from the proposed electromechanical tuning mechanism.

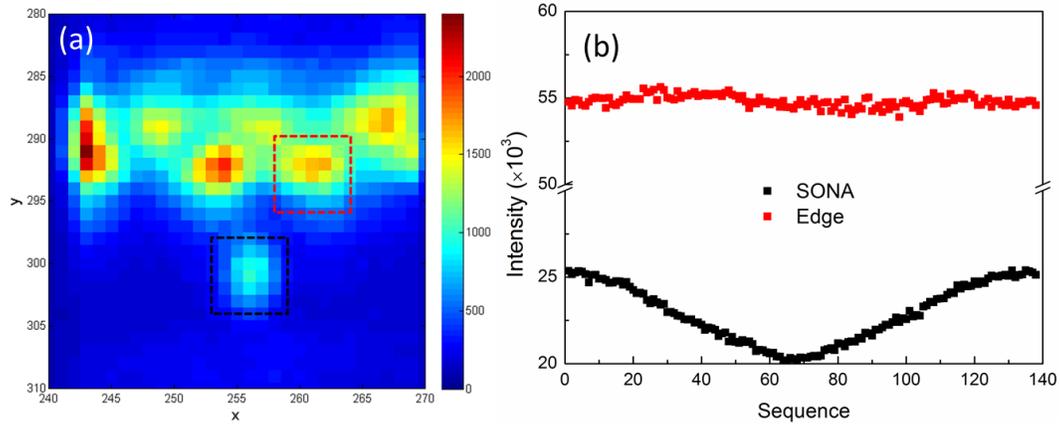

**Figure S4**. (a) Scattered light image of the antenna (black box) and the trench edge formed on the CCD camera of the spectrometer; (b) Changes of the scattered light intensity from the antenna and the reference point (red box in (a)). A break is used on the *y*-axis for better comparision of the data. The number of frame (or image) is recorded on *x*-axis.